\documentclass[a4paper,12pt]{article}
\usepackage[english,finnish,german]{babel}
\usepackage[paper=a4paper,left=20mm,width=170mm,top=30mm,bottom=25mm]{geometry}
\usepackage{setspace} % This is used in the title page
\usepackage{t1enc}
\usepackage{alltt}
\usepackage{epsfig}
\usepackage{graphicx,ifthen}
\usepackage{eurosym}
\usepackage{amsfonts}
\usepackage{latexsym}
\usepackage{amssymb,amsthm,amsmath}
\usepackage{verbatim}
\usepackage{multirow}
\usepackage{bm}
\usepackage{algorithm}
\usepackage{algorithmic}
\usepackage[T1]{fontenc}
\usepackage{color}
\usepackage{ae,aecompl}
\usepackage{indentfirst}
\begin{document}
\newpage

%\ttfamily
\selectlanguage{finnish}
% Set up a title page
\thispagestyle{empty} % no page number on very first page
% Use roman numerals for page numbers initially
\renewcommand{\thepage}{\roman{page}}

\begin{spacing}{1}
\begin{center}
{
%\vspace*{30mm}

%\includegraphics[width=5cm]{MonashCrest}

%\vspace*{15mm}

%\vspace*{10mm}

{\LARGE \bfseries Energy Efficient Resource Allocation and User Scheduling for Collaborative Mobile Clouds with Hybrid Receivers}

\vspace*{5mm}

{  Zheng Chang$^1$, \emph{Member, IEEE}, Jie Gong$^2$, \emph{Member, IEEE}, Tapani Ristaniemi$^1$,\emph{Senior Member, IEEE}, Zhisheng Niu$^2,\emph{Fellow, IEEE}$\\
 %%\vspace*{5mm}
$^1$Department of Mathematical Information Technology, University of Jyväskylä, \\P.O.Box 35(Agora), FIN-40014 Jyväskylä, Finland,
 \\e-mail: \{zheng.chang, tapani.ristaniemi \}@jyu.fi

$^2$ Department of Electronic Engineering, Tsinghua National Laboratory for Information Science and Technology, Tsinghua University, Beijing 100084, China
 \\e-mail: gongj13@mail.tsinghua.edu.cn, niuzhs@tsinghua.edu.cn

}
}

\end{center}
\end{spacing}
%\clearpage
\selectlanguage{english}

% Now reset page number counter,and switch to arabic numerals for remaining
% page numbers
\setcounter{page}{1}
\renewcommand{\thepage}{\arabic{page}}

\begin{abstract}
In this paper, we study the resource allocation and user scheduling algorithm for minimizing the energy cost of data transmission in the context of OFDMA collaborative mobile cloud (CMC) with simultaneous wireless information and power transfer (SWIPT) receivers. The CMC, which consists of several collaborating MTs offers one potential solution for downlink content distribution and for the energy consumption reduction at the terminal side. Meanwhile, as RF signal can carry both information and energy simultaneously, the induced SWIPT has gained much attention for energy efficiency design of mobile nodes. Previous work on the design of CMC system mainly focused on the cloud formulation or energy efficiency investigation, while how to allocate the radio resource and schedule user transmission lacks attention. With the objective to minimize the system energy consumption, an optimization problem which jointly considers subchannel assignment, power allocation and user scheduling for a group of SWIPT receivers has been presented. The formulated problem is addressed through the convex optimization technique. Simulation results demonstrate that the proposed user scheduling and resource allocation algorithms can achieve significant energy saving performance.
\end{abstract}


\section{Introduction}
\label{Sec1}

The conventional wireless networks are energy-constrained, in the sense that the network elements, such as mobile phones, are equipped with batteries. The battery capacity is improved at relatively slow speed during last decades, which creates the bottleneck in prolonging the lifetime of the networks. Meanwhile, the rise of online services significantly increases the demand for high data rate transmissions, hence straining the current network as well as drawing battery of mobile devices much faster than before. Therefore, novel energy-efficient approaches for network design are essential. Recently, there are increased interests on the energy-efficient design for the wireless communication systems. So called green communications, including the design of energy-efficient communication infrastructures, protocols, devices, and services, become inevitable trends for the evolution of the future networks. \par

Conventional studies on energy-constrained networks usually focused on designing the energy saving scheme proposals. Recently, energy harvesting (EH) technique received considerable attention due to its capability of realizing self-maintenance in wireless networks. Meanwhile, researches on EH have mainly concentrated on the transmitter side \cite{yang}. That is, for a cellular network, the recent work mainly focused on the development of BS to utilize the harvested energy rather than the mobile nodes, e.g., in \cite{Ho1}\cite{gong}. Although, there are many EH resources, such as solar, wind and tide, they are usually either location- or weather-dependent. Thus, for the indoor users who can not access solar or wind, EH becomes luxury or even impossible. These constraints motivate the wireless power transfer technology for the terminals, which enables the radio receiver to capture the RF radiation and convert into a direct current voltage \cite{Varshney}. \par

As RF signal can carry both information and energy simultaneously, the induced simultaneous wireless information and power transfer (SWIPT) has gained much attention \cite{Grover} \cite{huang}. Through SWIPT, the receiver not only can receive data from the transmitter, but also can "recycle" the transmit power to prolong the battery lifetime. Nowadays, electromagnetic wave is almost everywhere all around people's life, so enabling SWIPT is full of possibility in the future wireless networks. In \cite{Varshney} and \cite{Grover}, the fundamental trade-offs between wireless information and power transfer were studied with the assumption that the receiver can simultaneously receive information and harvest energy from the RF signal. In \cite{RZhang} and \cite{XZhou}, the authors proposed a receiver architecture which can split the received power into two power streams to facility the SWIPT. The authors of \cite{RZhang} and \cite{XZhou} also investigated the rate-energy regions for a SWIPT receiver in a two-user P2P scenario. The authors of \cite{Liu} focused on the optimization problem of power control and scheduling for SWIPT receiver. The optimal information decoding (ID) and EH mode switching was then derived. The work in \cite{ding} studied the outage probability of a cooperative EH relay network with multiple transmission pairs. In \cite{ng} and \cite{ng2}, different subcarrier and power allocation algorithms were proposed for the multiuser OFDM systems with SWIPT. Non-convex optimization problem was formulated with the objective to maximize the energy efficiency performance in term of $bits/Joule$. For a large scale MIMO system with SWIPT receivers, the authors of \cite{chen} presented an optimization scheme to maximize the energy efficiency of the system while satisfying the delay constraints by jointly allocating the transmission time duration and transmit power. \par

Meanwhile, the multimedia services, such as news download (e.g., breaking news), multimedia multicasting (e.g., live sport events or videos), or content distribution (e.g., device configuration files/pictures), are facing increasingly popularity in the daily life. Hence, how to efficiently provide those services for various users also attracts many interests. One of the potential solutions for content distribution in the mobile networks is to design a cooperative content distribution architecture named Collaborative Mobile Cloud (CMC) where the users can share some content and information cooperatively through Device-to-Device (D2D) or Machine-to-Machine (M2M) communications \cite{pederson}. In addition, CMC is foreseeable to reduce the energy consumption of mobile terminals (MTs) as well \cite{chang1}. Different from the concepts of cloud computing and cloud radio access network (CRAN) where a centralized entity with rich computational resources executes the computing or data processing tasks instead of MT, CMC is formed by a number of MTs who can process data in a distritbuted and cooperative manner. In the CMC, the MTs can actively use two wireless interfaces: one to communicate with the Base Station (BS) over a Long-Range (LR) wireless technology (such as LTE) and the other to cooperate with other MTs over a Short-Range (SR) communication link (such as WLAN/ac hoc) \cite{alkanj}. In the traditional service, each MT has to download the whole content on its own, which leads to significant energy consumption, especially if the LR data rate is low. In the CMC system, several MTs can form a coalition and each MT receives parts of required information data from BS. Then they exchange the received data with others \cite{chang1}. In such case, each MT only needs to download parts of the data and consequently, the receiving time can be significantly reduced. Although information exchange over SR introduces new transmission overhead, the energy consumption can still be significantly decreased as the SR is generally more efficient in term of data rate \cite{lina}.

As SWIPT enables the traditional information receiver to harvest energy from transmission signal, a CMC formed by a group of SWIPT receivers is expected to improve energy saving performance. Previous work on the design of CMC mainly focused on the cloud formulation or energy efficiency investigations, how to allocate the limited radio resources such as subchannels or transmit power lacks concerns. However, as the future wireless network is OFDMA-based, efficient and practical radio resource allocation scheme is critical. Intuitively, there are two ways for delivering data between BS and CMC. One is that BS can transmit different packets to different UEs in a parallel way. In this case, more channel bandwidths are needed as BS is transmitting to different UEs with different data simultaneously. The other one is that BS can transmit to the UEs in a sequence. For each data segment, one UE is selected as the receiver for BS and transmitter for other UEs. As such, channel bandwidth can be effectively utilized. In this work, we focus on the latter case. On the way towards resource allocation for CMC with SWIPT receivers, there are some obstacles:

\begin{itemize}
  \item How to schedule the proper MT to receive from BS and transmit to other MTs is essential. As CMC is expected as an energy-efficient content distribution system, the MTs being selected should be able to minimize the overall system energy cost;
  \item How to properly assign the subchannels for the transmission between BS and CMC and the transmission inside CMC should also be concentrated. Moreover, power allocation schemes for both BS and MTs need to be investigated so that the transmit power consumption can be minimized;
  \item The ID and EH functionalities of receivers make the problem even more complicated. At each scheduling time, which MT should be selected so that the harvested energy can be maximized while the receive energy consumption can be minimized? In addition, since eventually better channel condition and higher transmit power of BS can bring more energy for harvesting, which set of subchannnels and how much transmit power should be used so that the sum of the consumed energy and harvested energy can be minimized call for contiguous consideration.
\end{itemize}

In this work, we address the user scheduling and resource allocation problem for the considered system. In order to utilize the channel bandwidth, during the BS data delivery process, one dedicated receiver, denoted as information decoding MT (IMT), will receive the assigned data and the other MTs, denoted as energy harvesting MTs (EMTs), will harvest energy from the signal. As such, the transmission of the system is in a sequence and no extra bandwidth other than the conventional multicasting is required. After receiving from BS, IMT will share the data to other EMTs of the previous stage. Solving the formulated problem is challenging due to the aforementioned obstacles. To address the formulated the problem, different nonlinear programming methods are applied. The main contribution over existing works is three-fold:

\begin{itemize}
  \item We first model the energy consumption of the overall transmission process when considering baseband circuit energy consumption, RF transmit and receive energy consumption and harvested energy;
  \item Then we focus on the algorithm design aspect and propose a joint power allocation, subchannel allocation and user scheduling scheme with the objective of optimizing energy consumption performance of CMC;
  \item By using nonlinear fractional programming optimization techniques and iterative algorithm design, we address the formulated user scheduling and resource allocation problem. Simulation results are presented to illustrate the energy saving gain of the proposed scheme.
\end{itemize}

The rest of this paper is organized as follows. Section \uppercase\expandafter{\romannumeral2} describes the collaborative mobile cloud system model. In Section \uppercase\expandafter{\romannumeral3}, energy consumption models of ID and EH receivers, as well as CMC are presented. In Section \uppercase\expandafter{\romannumeral4}, we formulate the optimization problem and introduce resource allocation and user scheduling solution. We demonstrate the benefits of our proposed algorithm in Section \uppercase\expandafter{\romannumeral5} through simulation study and finally conclude this work in Section \uppercase\expandafter{\romannumeral6}.

\section{System Model}
\label{Sec2}

In our considered OFDMA system, it is assumed that $K$ MTs locating geographically close are interested in downloading the same content from a BS using one LR wireless technology (e.g., UMTS/HSPA, WiMAX, or LTE). In the CMC system, a MT firstly receives parts of the data and then shares with each other through SR transmission via D2D/M2M connections \cite{chang1} so that the receiving time of other MTs can be reduced. The MTs are able to decode information and harvest energy from the received radio signals. All transceivers are equipped with a single antenna. Moreover, We assume the channel follows quasi-static block fading and the channel state information can be accurately obtained by the feedback from the receivers. \par

\begin{figure}[t]
\centering
\includegraphics[height=6cm, width=8cm]{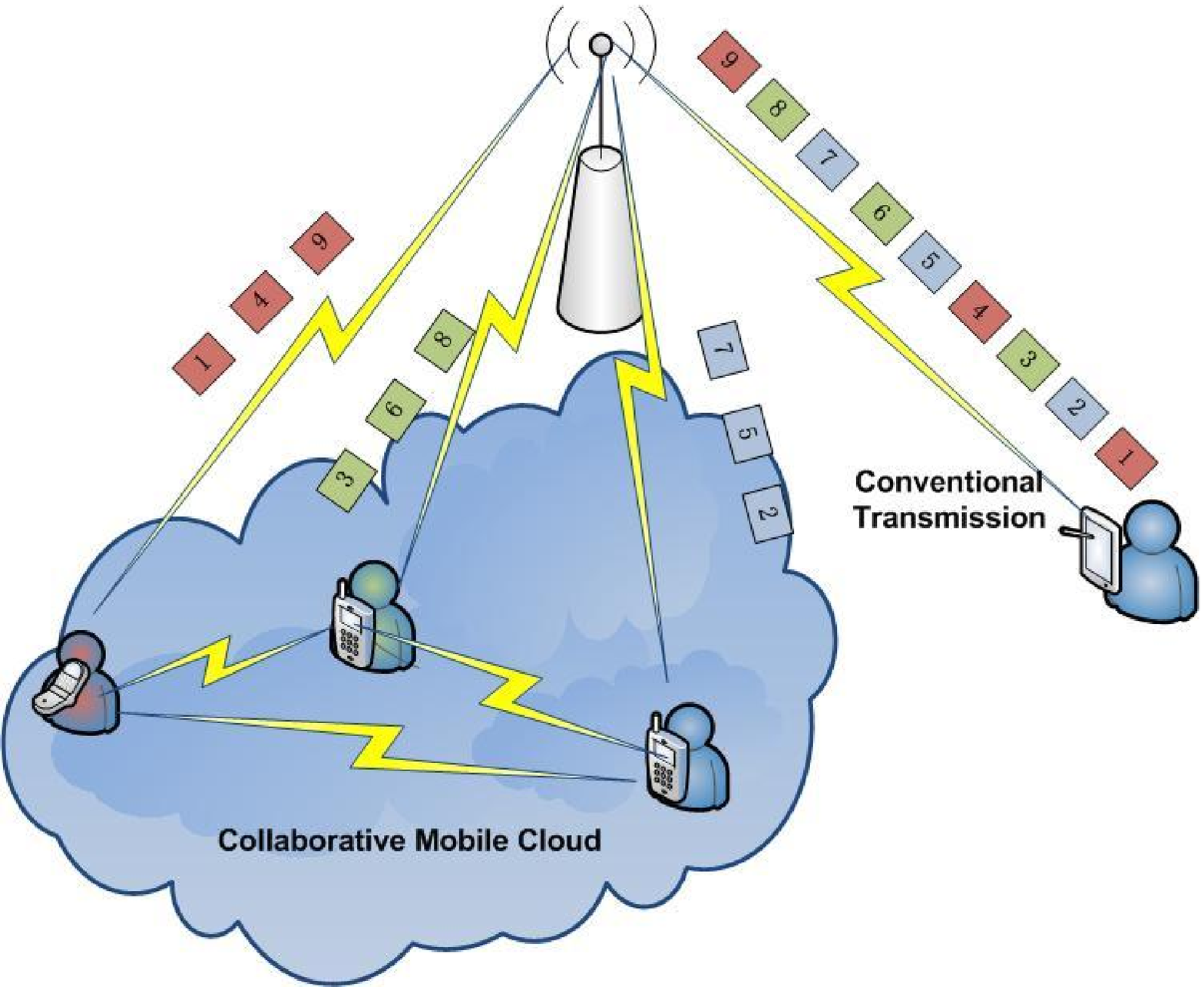}
\caption{Transmission Procedure} \label{fig:1}
\end{figure}

\begin{figure}[t]
\centering
\includegraphics[height=8cm, width=11cm]{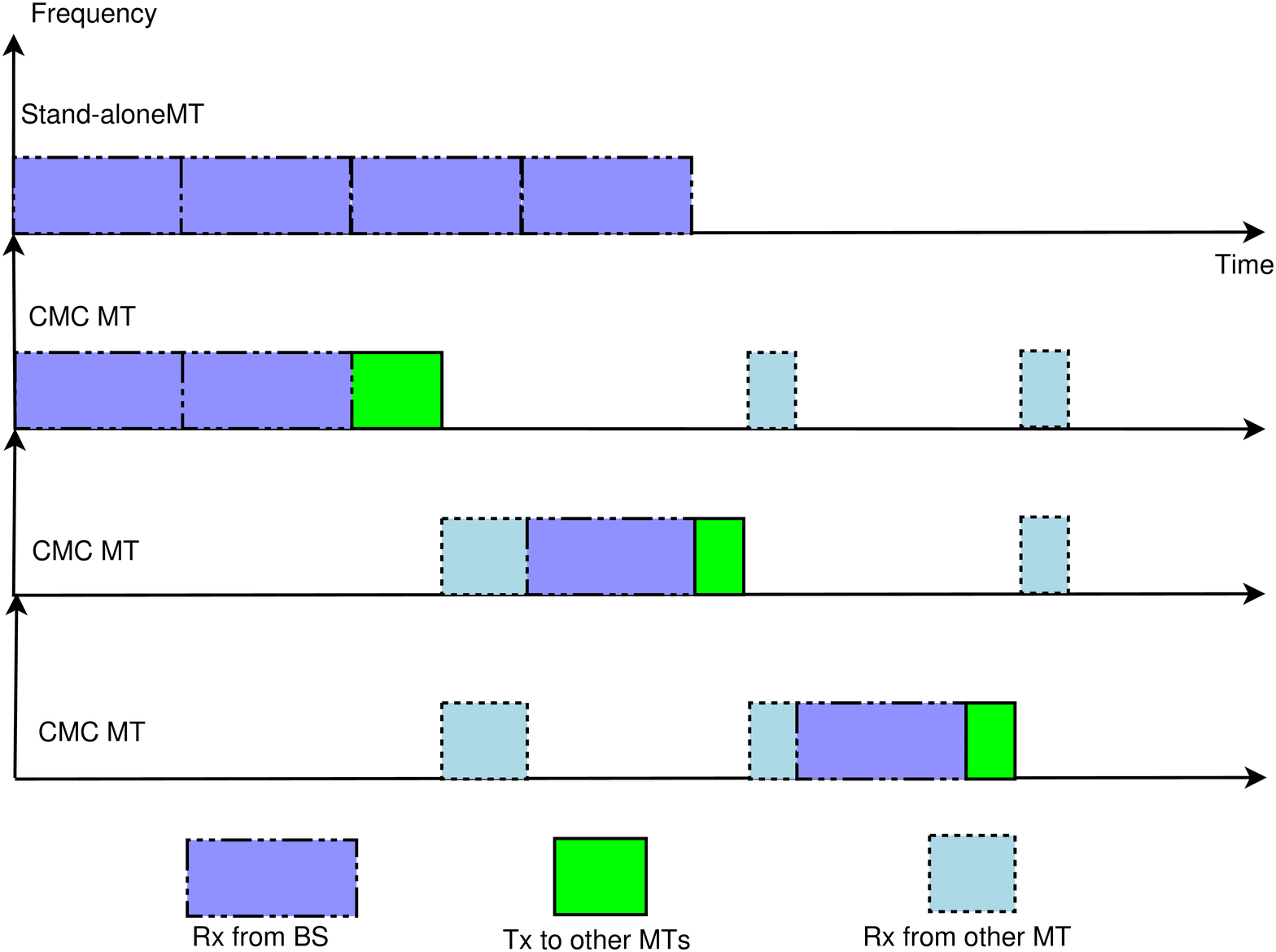}
\caption{Transmission Procedure} \label{fig:12}
\end{figure}

The transmission process of CMC is depicted in Fig. \ref{fig:1} comparing with the conventional transmission, e.g., unicasting/multicasting. In Fig. \ref{fig:1}, the CMC consists of three MTs. To simply present the concept of the CMC, we divide the overall data stream into $9$ segments. In a conventional setup,  the communication interface of a MT (e.g. stand-alone MT) has to remain active for the whole reception duration. This results in high energy consumption due to the required RF and baseband processing during data reception. In the CMC, BS can distribute various and exclusive segments of data (3 segments in the figure) to different MTs, and MTs can then exchange/share the received parts with other coalition members by utilizing the SR transmission among MTs. Thus, the reception duration can be seriously reduced compared to the conventional transmission given that SR has better a data rate than LR because of, e.g., its better channel condition. The time sequence of the transmission process is shown in Fig. \ref{fig:12} where how the transmission duration can be reduced can be found. However, in order to share the received data, additional communication overheads are induced, such as transmit power of SR transmitter, transmit/receive durations. Consequently, the inherent energy efficiency performance calls for a careful algorithm design \cite{chang2}.

%When considering the CMC with hybrid ID-EH receivers, the CMC members are able to receive power transfer when not being assigned data. Indeed, the SWIPT provides one potential solution for stimulating users to join and form the CMC. In the following, we will present the model for SWIPT receiver as well as formulate the resource allocation and user scheduling optimization problem in a CMC with hybrid receivers.\par

In order to focus on the resource allocation algorithm design and isolate it from the specific hardware implementation details, we do not assume a particular type of EH receiver. In this work, we focus on the receivers which consist of an energy harvesting unit and a conventional signal processing core unit for energy harvesting and information decoding, respectively. In addition, for the conventional signal processing, we separate receiver architecture into two parts, which are RF unit and baseband unit. In the following section, the energy consumption model of the considered system is presented. \par

\section{Energy Consumption Model}

\subsection{Energy Harvesting Receiver}
\label{Sec2.2}

Let $x$ be the transmitted data from BS, the received data can be modeled as

\begin{equation}
\label{eq:basicReceivedSignal}
 y = \sqrt{P_{s,k,i}^{L_{tx}} L_{s,k} H_{s,k,i}} x + z_{s,k,i},
\end{equation}

\noindent where $P_{s,k,i}^{L_{tx}}$ and $H_{s,k,i}$ are the transmit power and channel fading gain on subchannel $i$ from BS to MT $k$, respectively. $L_{s,k}$ represents the path loss from BS to MT $k$. $z_{s,k,i}$ is the Gaussian additive noise with zero mean and variance $\sigma_{z}^2$.\par

In practice, the model of an EH receiver depends on its specific implementation. For example, electromagnetic induction and electromagnetic radiation are able to transfer wireless power, and the receiver is able to recycle the wireless power from radio signal \cite{Grover}. Nevertheless, the associated hardware circuit in the receivers and the corresponding EH efficiency can be different. Besides, the signal used for decoding the modulated information cannot be used for harvesting energy due to hardware limitations \cite{RZhang}. In this work, we do not go into the details of any particular type of EH receiver and the results of \cite{XZhou} are utilized. \par

We denote $P_{s,k,i}^H$ as the harvested power on subchannel $i$ by EMT $k$. Then we have \cite{XZhou}

\begin{equation}
\label{eq:harvestEnergy}
P_{s,k,i}^H =  \vartheta_{k} P_{s,k,i}^{L_{tx}} L_{s,k} H_{s,k,i},
\end{equation}

\noindent where we assume the conversion efficiency $0 < \vartheta_{k} \leq 1$.

\subsection{Information Decoding Receiver}
\label{Sec2.3}

For the data rate on LR $R_{s,k,i}^{L}$, the maximum achievable data rate in $bps/Hz$ from BS to MT $k$ on subchannel $i$ is given as

\begin{equation}
\label{eq:Capacity}
R_{s,k,i}^L =  \log_2 \left(1+\frac{P_{s,k,i}^{L_{tx}} L_{s,k} H_{s,k,i}}{\sigma_z^2} \right).
\end{equation}

In the CMC, the IMT $k$ needs to multicast its received data to other CMC members, so the data rate on the subchannel $j$ of SR link can be expressed as

\begin{equation}
\label{eq:Capacity2}
R_{k,j}^{S} = \log_2 \left(1+\frac{P_{k,j}^{S_{tx}}L_{k} H_{k,j}}{\sigma_z^2} \right),
\end{equation}

\noindent where $P_{k,j}^{S_{tx}}$ is the multicast transmit power of IMT $k$, $L_{k}$ and $H_{k,j}$ are the path loss and channel gain from $k$ to the MT with worst channel condition, respectively. We also assume that the noises on LR and SR are of same kind for simplicity.

\subsection{Tx and Rx Energy Consumption of CMC}
\label{Sec2.4}

As we know, the energy consumption can be modeled as a linear function containing the power consumption and the time duration. Therefore, the energy consumption for receiving data size $S_T$ from BS can be expressed as

\begin{equation}
\label{eq:LRRX}
\begin{split}
E_{s,k,i}^{L_{rx}}  & = (P_{s,k,i}^{L_{rx}} + P_{E}) T_{s,k,i}^{L_{rx}}  = \frac{(P_{s,k,i}^{L_{rx}} + P_{E}) S_T}{R_{s,k,i}^{L}} \\ &= \frac{(P_{rx}+ P_{E})S_T}{R_{s,k,i}^{L}},
\end{split}
\end{equation}

\noindent where $P_{s,k,i}^{L_{rx}}$ is the RF power consumption of $k$ for receiving from BS on subchannel $i$ and $P_{E}$ is the electronic circuit power consumption of baseband associated with transmission bandwidth. In this work, the energy consumption refers to the one when receiving and sending data on certain subchannel, so the baseband power consumption is considered together with RF Tx/Rx power consumption. $T_{s,k,i}^{L_{rx}} = \frac{S_T}{R_{s,k,i}^{L}}$ is the required time for receiving data $S_T$ on LR subchannel $i$. Further we can assume the receive RF power consumption are the same for both LR and SR links, and equals to $P_{rx}$. After receiving from the BS, MT $k$ is going to transmit its received data to other required MTs. There are two conventional ways to deliver data inside CMC, which are unicasting and multicasting. We have discussed the energy efficiency of using both two schemes in \cite{chang2} and the multicasting shown the superior energy efficiency performance over unicasting. So in this work, we only invoke multicasting as the transmission strategy inside CMC. \par

When multicasting is used, an IMT only needs to broadcast its data to other MTs in CMC once with the data rate that can reach the MT with the worst channel condition. Thus the transmit energy consumption of IMT is given as

\begin{equation}
\label{eq:SRmulticastTX}
E_{k,j}^{S_{tx}} = (P_{k,j}^{S_{tx}}+ P_{E}) T_{k,j}^{S_{tx}} = \frac{(P_{k,j}^{S_{tx}}+ P_{E})
S_T}{R_{k,j}^{S}}.
\end{equation}

Therefore, the total energy consumption of CMC when using MT $k$ as the IMT can be expressed as follows:

\begin{equation}
\label{eq:SRmulticastMT}
E_{k,i,j} = E_{s,k,i}^{L_{rx}}+E_{k,j}^{S_{tx}}+ \sum_{n,n \neq k}^K E_{n,j}^{S_{rx}}.
\end{equation}

\noindent $E_{k,i,j}$ is the energy consumption of IMT $k$ when assigning subchannel $i$ for receiving from BS and subchannel $j$ for broadcasting its received data. $E_{n,j}^{S_{rx}}$ is the energy consumption of each EMT when receiving from IMT on subchannel $j$, and it can be expressed as

\begin{equation}
\label{eq:SRmulticastRX}
\begin{split}
E_{n,j}^{S_{rx}} &= (P_{n,j}^{S_{rx}}+ P_{E}) T_{k,j}^{S_{rx}} = \frac{(P_{n,j}^{S_{rx}}+ P_{E})
S_T}{R_{k,j}^{S}} \\&= \frac{(P_{rx}+ P_{E})S_T}{R_{k,j}^{S}}.
\end{split}
\end{equation}

\subsection{Energy Consumption of Base Station}
\label{Sec2.5}

The energy consumption of the BS is given as

\begin{equation}
\label{eq:LRTX}
E_{s}^{L_{tx}} = (P_{s,k,i}^{L_{tx}} + P_{B}) T_{s,k,i}^{L_{rx}} = \frac{(P_{s,k,i}^{L_{tx}} + P_{B}) S_T}{R_{s,k,i}^{L}},
\end{equation}

\noindent where $P_B$ is the BS baseband operating power consumption. \par

\section{Resource Allocation and User Scheduling}
\label{Sec3}

\subsection{Problem Formulation}
\label{Sec3.1}

In the previous section, we presented the energy consumption model of the CMC with hybrid ID-EH receivers in each scheduling time slot. In this section, the considered problem will be formulated as a joint optimization of subchannel allocation, power allocation and user scheduling with the objective to minimize the energy consumption during each transmission time of sending data $S_T$. \par

In order to minimize the energy consumption of a CMC, at first one MT inside CMC will be selected as the IMT and other MTs will be considered as EH receivers. Then the BS transmits the data to the IMT on the selected subchannel $i$ with data rate $R_{s,k,i}^{L}$. After that, the IMT will act as a relay and forwards the received data to other MTs on the selected subchannel $j$ with multicast data rate $R_{k,j}^{S}$. In this work, we tackle the problem when one subchannel is used for LR link and one is used for SR link. However, the work can be applied to multiple-subchannel problems as well.  \par

%under different circumstance, such as traditional multicasting consists of multiple channel bandwidths.\par

To this end, taken harvested energy into consideration, for each data segment transmission, we can formulate the optimization objective as

\begin{equation}
\begin{split}
\label{eq:totalenergy}
\mathcal{E} (\bm{\rho},\bm{\omega}, \mathbf{P})& = \sum_{k=1}^{K} \sum_{i=1}^{N} \rho_k \omega_{s,k,i} E_{s}^{L_{tx}} \\ & +  \sum_{k=1}^{K} \sum_{i=1}^{N} \sum_{j=1}^{N} \omega_{s,k,i} \omega_{k,j} \rho_k  E_{k,i,j} \\& - \sum_{i=1}^{N} \sum_{k=1}^{K}  \sum_{n,n \neq k}  \omega_{s,k,i} \rho_k  Q_{s,n,i},
\end{split}
\end{equation}

\noindent where $N$ is the number of available subchannels, and $K$ is the total number of MTs inside CMC. $\mathbf{P}$ is the power allocation policy. $\bm{\rho} =\{ \rho_k\}, \forall k $ and $\bm{\omega} = \{ \omega_{s,k,i}, \omega_{k,j} \}, \forall k,i,j $ are the user selection and subchannel allocation indicators. $Q_{s,n,i}$ is the harvest energy and obversely  $Q_{s,n,i} = P_{s,n,i}^H S_T/R_{s,k,i}^{L}$. Note that mathematically $\mathcal{E} (\bm{\rho},\bm{\omega}, \mathbf{P})$ can take the negative value. However, the case that $\mathcal{E} (\bm{\rho},\bm{\omega}, \mathbf{P})$ is positive always holds for practical system. The defined binary variable $\rho_{k}$ is the indicator whether MT $k$ is selected as IMT, that is,

\begin{equation}
 \rho_{k}=
\begin{cases}
1, &\text{if $k$ is chosen as IMT for receiving from BS,}\\
0, &\text{otherwise.}
\end{cases}
\end{equation}

\noindent In addition, we also define $\omega$ is the indicator whether certain subchannel is assigned to MT $k$, e.g.,

\begin{equation}
 \omega_{s,k,i}=
\begin{cases}
1, &\text{if subchannel $i$ is used by $k$ for downlink,}\\
0, &\text{otherwise.}
\end{cases}
\end{equation}

\noindent and

\begin{equation}
 \omega_{k,j}=
\begin{cases}
1, &\text{if subchannel $j$ is assigned to $k$ to delivery data}\\
0, &\text{otherwise.}
\end{cases}
\end{equation}

Therefore, the user selection and resource allocation optimization problem can be formulated as

\begin{equation}
\label{eq:energyOP}
\min_{\bm{\rho},\bm{\omega}, \mathbf{P}}\mathcal{E} (\bm{\rho},\bm{\omega}, \mathbf{P}),
\end{equation}

\noindent s.t.

\begin{equation}
\label{eq:energyOPconstraints}
\begin{split}
C1:\ & \sum_{k=1}^{K} \rho_k  = 1,\\
C2:\ & \sum_{k=1}^{K} \omega_{s,k,i}  = 1,  \sum_{k=1}^{K} \omega_{k,j}  = 1, \\
C3:\ & R_{s,k,i}^{L} \geq R_{L,min}, \\
C4:\ & R_{k,j}^{S} \geq R_{S,min},\\
C5:\ & \sum_{k=1}^{K} \sum_{i=1}^{N} \rho_k \omega_{s,k,i} P_{s,k,i}^{L_{tx}} \leq P_{s,max}, \\
C6:\ & \sum_{k=1}^{K} \sum_{j=1}^{N} \rho_k \omega_{k,j} P_{k,j}^{S_{tx}} \leq P_{k,max},\\
%C7:\ & \sum_{k=1}^{K} \rho_k\! \left(\! E_{k,re} \! -\! \sum_{i=1}^{N}\sum_{j=1}^{N}\omega_{s,k,i}\omega_{k,j} (E_{s,k,i}^{L_{rx}}\!+\!E_{k,j}^{S_{tx}})\! \right)\!\geq\! E_{ma},
\end{split}
\end{equation}

Here, the optimization problem (\ref{eq:energyOP}) is formulated with several constraints. The first constraint C1 is to ensure only one user is selected for receiving data from BS in a scheduled slot. C2 ensures the subchannels allocated to MT $k$ are unique. $R_{L,min}$ in C3 and $R_{S,min}$ in C4 are the required data rates for LR and SR, respectively. C5 and C6 ensure that the power allocation of BS and IMT should not be higher than the maximum allowed transmit power. \par

%C7 is imposed to guarantee that if MT $k$ is selected as IMT, $k$ still should have the maintenance energy $E_{ma}$ after delivering data to others.
It is worth noticing that (\ref{eq:energyOP}) with (\ref{eq:energyOPconstraints}) is combinatorial in nature with a non-convex structure. In general, there is no standard approach for solving such a non-convex optimization problems and such integer programming problem is recognized as NP-hard. In the extreme case, an exhaustive search or branch-and-bound method is needed to obtain the global optimal solution which requires high computational complexity even for small $K$ and $N$. In order to make the problem tractable, we transform the objective function and approximate the transformed objective function in order to simplify the problem.

\subsection{Power and Subchannel Allocation Scheme}
\label{Sec3.2}

%\begin{equation}
%\begin{split}
%\label{eq:totalenergyRef}
%\mathcal{E} (\bm{\rho},\bm{\omega}, \mathbf{P}) &  =
% \frac{\sum_{k=1}^{K} \rho_k S_T \left (P_{s,k,i}^{L_{tx}} + P_{B} + \sum_{i=1}^{N} \sum_{j=1}^{N} \omega_{s,k,i} \omega_{k,j} (P_{rx}+ P_{E})- \sum_{i=1}^{N} \sum_{n,n \neq k} \omega_{s,k,i} \vartheta_{n} P_{s,k,i}^{L_{tx}} L_{s,n} H_{s,n,i} \right)}{R_{s,k,i}^{L}}  \\&+ \frac{\sum_{k=1}^{K} \sum_{i=1}^{N} \sum_{j=1}^{N} \omega_{s,k,i} \omega_{k,j} \rho_k  S_T(P_{k,j}^{S_{tx}}+  P_{rx}+ 2P_{E})}{R_{k,j}^{S}}
%\end{split}
%\end{equation}%
%\end{split}
%\end{equation}

\newtheorem{theorem}{Theorem}
\begin{theorem}
The objective function (\ref{eq:energyOP}) is quasi-convex function w.r.t.e to the power allocation policy.
\end{theorem}

The proof of \textbf{Theorem 1} can be found in Appendix A. Thus, as a result, an unique global optimal solution exists and the optimal point can be obtained by using the bisection method \cite{boyd}. We can also apply the nonlinear fractional programming method to solve the formulated problem \cite{Dinkelbach} of power allocation and subchannel allocation in the followings. \par

\subsubsection{Problem Transformation}
\label{Sec3.2.1}

First given that the user scheduling is done, i.e. $\rho_{k} = 1$, we can reform the objective function $\mathcal{E} (\bm{\rho},\bm{\omega}, \mathbf{P})$ as a function of $\{\bm{\omega},\mathbf{P}\}$. Substituting (\ref{eq:harvestEnergy}), (\ref{eq:SRmulticastMT}) and (\ref{eq:LRTX}) into (\ref{eq:totalenergy}), we can arrive (\ref{eq:totalenergyRef}) where $P_c  = P_{B} +  P_{rx}+ P_{E}$. One may notice that obtaining power allocation policy involves solving $\mathcal{E} (\bm{\omega},\mathbf{P})$, which can be expressed as

%---------------------this is used for put eq to the beginning of the page--------------------------------
%\newcounter{MYtempeqncnt}
%\begin{figure*}[t]
%\normalsize
%\setcounter{MYtempeqncnt}{\value{equation}}
%\setcounter{equation}{15}
\begin{equation}
\begin{split}
\label{eq:totalenergyRef}
\mathcal{E} (\bm{\omega},\mathbf{P})  = &
 \frac{\overbrace {S_T \left (  \sum_{i=1}^{N} \omega_{s,k,i} P_{s,k,i}^{L_{tx}} + P_{c}-  \sum_{i=1}^{N} \sum_{n,n \neq k} \omega_{s,k,i} \vartheta_{n} P_{s,k,i}^{L_{tx}} L_{s,n} H_{s,n,i} \right)}^{U_1(\omega_{s,k,i}, P_{s,k,i}^{L_{tx}})}}{\underbrace{\sum_{i=1}^{N} \omega_{s,k,i} R_{s,k,i}^{L}}_{R_1(\omega_{s,k,i},P_{s,k,i}^{L_{tx}})}} \\+& \frac{ \overbrace{S_T\sum_{j=1}^{N} \omega_{k,j}(P_{k,j}^{S_{tx}}+ (K-1) P_{rx}+ KP_{E})}^{U_2(\omega_{k,j},P_{k,j}^{S_{tx}})}}{ \underbrace{\sum_{j=1}^{N}\omega_{k,j}R_{k,j}^{S}}_{R_2(\omega_{k,j},P_{k,j}^{S_{tx}})}}.
\end{split}
\end{equation}%

%\setcounter{equation}{\value{MYtempeqncnt}}
%\hrulefill
%\vspace*{4pt}
%\end{figure*}
%---------------------------------------------------------------------------------------------------

%\setcounter{equation}{16}
\begin{equation}
\label{eq:totalenergyRef2}
\mathcal{E} (\bm{\omega}, \mathbf{P}) = \mathcal{E}_{LR}(\omega_{s,k,i},P_{s,k,i}^{L_{tx}}) + \mathcal{E}_{SR}(\omega_{k,j},P_{k,j}^{S_{tx}}),
\end{equation}

\noindent where $\mathcal{E}_{LR}(\omega_{s,k,i},P_{s,k,i}^{L_{tx}}) = \frac{U_1(\omega_{s,k,i}, P_{s,k,i}^{L_{tx}})}{R_1(\omega_{s,k,i}, P_{s,k,i}^{L_{tx}})}$ and $\mathcal{E}_{SR}(\omega_{k,j}, P_{k,j}^{S_{tx}})=  \frac{U_2(\omega_{k,j}, P_{k,j}^{S_{tx}})}{R_2(\omega_{k,j}, P_{k,j}^{S_{tx}})}$. From (\ref{eq:totalenergyRef2}), one can observe that the power allocation schemes for BS and scheduled MT are separated. In other word, we can obtain optimal power allocation by addressing $\mathcal{E}_{LR}(\omega_{s,k,i},P_{s,k,i}^{L_{tx}})$ and $\mathcal{E}_{SR}(\omega_{k,j},P_{k,j}^{S_{tx}})$ individually when user scheduling is done. We can see that both  $\mathcal{E}_{LR}(\omega_{s,k,i},P_{s,k,i}^{L_{tx}})$ and $\mathcal{E}_{SR}(\omega_{k,j},P_{k,j}^{S_{tx}})$  are quasi-convex functions w.r.t. power allocation variables. For the sake of presentation simplicity, we introduce a method for solving $\mathcal{E}_{LR}(\omega_{s,k,i},P_{s,k,i}^{L_{tx}})$ which is derived from nonlinear fractional programming \cite{Dinkelbach}. \par

The global optimal solution $q_{LR}^*$ can be expressed as

\begin{equation}
\label{eq:optimalq}
q_{LR}^* = \mathcal{E}_{LR}(\omega_{s,k,i}^*,P_{s,k,i}^{L*})=  \min_{\omega_{s,k,i},P_{s,k,i}^{L_{tx}}} \frac{U_1(\omega_{s,k,i},P_{s,k,i}^{L_{tx}})}{R_1(\omega_{s,k,i},P_{s,k,i}^{L_{tx}})}.
\end{equation}

\begin{theorem}
The optimal solution $q_{LR}^*$ can be obtained iff
\begin{equation}
\label{eq:optimalqCon}
\min_{\omega_{s,k,i},P_{s,k,i}^{L_{tx}}} U_1(\omega_{s,k,i},P_{s,k,i}^{L_{tx}}) - q_{LR}^* R_1(\omega_{s,k,i},P_{s,k,i}^{L_{tx}}) = 0.
\end{equation}

\end{theorem}

\textbf{Theorem 2} gives a necessary and sufficient condition w.r.t. optimal power allocation. The proof can be found in Appendix B. Particularly, for the considered optimization problem with an objective function in fractional form, there exists an equivalent optimization problem with an objective function in subtractive form, i.e., $U_1(\omega_{s,k,i}, P_{s,k,i}^{L_{tx}}) - q_{LR}^* R_1(\omega_{s,k,i},P_{s,k,i}^{L_{tx}})$, and both formulations result in the same power allocations. To achieve the optimal $q_{LR}^*$, the iterative algorithm with guaranteed convergence in \cite{Dinkelbach} can be applied. The proof is given in Appendix C and the iterative algorithm is given in Alg. \ref{alg:IRA}.

\begin{algorithm}[htb]
\caption{Iterative Algorithm for Obtaining $q_{LR}^*$}
\label{alg:IRA}
\begin{algorithmic}[1]
\WHILE{(!Convergence)}
\STATE Solve the problem (\ref{eq:optimaNew}) for a given $q_{LR}$ and obtain subchannel and power allocation $\{\bm{\omega}^{\prime},\bm{P}^{\prime}\}$;\
\IF{$U_1(\bm{\omega}^{\prime},\bm{P}^{\prime})-q_{LR}R_1(\bm{\omega}^{\prime},\bm{P}^{\prime})=0$}
\STATE Convergence $=$ true; \
\RETURN $\{\bm{\omega}^{\ast},\bm{P}^{\ast}\} = \{\bm{\omega}^{\prime},\bm{P}^{\prime}\}$ and obtain $q_{LR}^*$ by (\ref{eq:optimalq});
\ELSE
\STATE Convergence $=$ false; \
\RETURN Obtain $q_{LR}= U_1(\bm{\omega}^{\prime},\bm{P}^{\prime})/R_1(\bm{\omega}^{\prime},\bm{P}^{\prime})$; \
\ENDIF
\ENDWHILE
\end{algorithmic}
\end{algorithm}

During the iteration, in order to achieve $q_{LR}^*$, we need to address the following problem with $q_{LR}$:
\begin{equation}
\label{eq:optimaNew}
\min_{\omega_{s,k,i},P_{s,k,i}^{L_{tx}}} U_1(\omega_{s,k,i},P_{s,k,i}^{L_{tx}}) - q_{LR} R_1(\omega_{s,k,i},P_{s,k,i}^{L_{tx}}),
\end{equation}

\noindent s.t.

\begin{equation}
\label{eq:energyOPconstraints2}
\begin{split}
\ & \sum_{k=1}^{K} \omega_{s,k,i}  = 1, \\
\ & R_{s,k,i}^{L} \geq R_{L,min}, \\
\ & \sum_{i=1}^{N} \omega_{s,k,i} P_{s,k,i}^{L_{tx}} \leq P_{s,max}, \\
%\ & E_{k,re} -\sum_{i=1}^{N}\sum_{j=1}^{N}\omega_{s,k,i}\omega_{k,j} (E_{s,k,i}^{L_{rx}}+E_{k,j}^{S_{tx}})  \geq E_{ma},
\end{split}
\end{equation}

Basically, such problem still is a non-convex optimization problem due to the involved integer programming. Tackling the mix convex and combinatorial optimization problem and obtaining a global optimal solution result in a prohibitively high complexity w.r.t. $K$ and $N$. Another solution which can balance the computational complexity and optimality can be obtained when addressing such problem in the dual domain. As discussed in \cite{wyu}, in the considered multi-carrier systems the duality gap of such a non-convex resource allocation problem satisfying the time-sharing condition is negligible as the number of subcarriers becomes sufficiently large e.g., 64. Since our optimization problem obviously is able to satisfy the time-sharing condition, it can be solved by using the dual method and the solution is asymptotically optimal. The same procedure can be used for achieving $\omega_{k,j}^*$ and $P_ {k,j}^{S_{tx}*}$. \par

\subsubsection{Dual Problem Formulation and Decomposition}
\label{Sec3.2.2}

In this part, we solve the resource allocation optimization problem of LR link by solving its dual for a given value of $q_{LR}$. The Lagrangian function of the primal problem (\ref{eq:optimalqCon}) can be given as,

\begin{equation}
\begin{split}
\label{eq:Lagrangian}
& \mathcal{L}(\mathbf{P},\bm{\omega},\lambda,\mu,\theta, \gamma) =
U_1(\omega_{s,k,i},P_{s,k,i}^{L_{tx}}) \\-& q_{LR}R_1(\omega_{s,k,i},P_{s,k,i}^{L_{tx}})-
\lambda(\sum_{k=1}^{K} \omega_{s,k,i} - 1)\\
-&\mu(R_{s,k,i}^{L} - R_{L,min} ) - \theta(P_{s,max} - P_{s,k,i}^{L_{tx}} ) ,\\
%- & \gamma\left( E_{k,re} -\sum_{i=1}^{N}\sum_{j=1}^{N}\omega_{s,k,i}\omega_{k,j} (E_{s,k,i}^{L_{rx}}+E_{k,j}^{S_{tx}} )- E_{ma}\right) ,\\
\end{split}
\end{equation}

\noindent where $\lambda,\mu,\theta$ are the lagrange multipliers associated with different constraints. Therefore, the dual problem is

\begin{equation}
\label{eq:maxdualfunction} \mathop{\max}_{\lambda,\mu,\theta} \ \min_{\mathbf{P},\bm{\omega}} \mathcal{L}(\mathbf{P},\bm{\omega},\lambda,\mu,\theta).
\end{equation}

By using Lagrange dual decomposition, the dual problem (\ref{eq:maxdualfunction}) can be decomposed into two layers, minimization of (\ref{eq:Lagrangian}) which is the inner problem and maximization of (\ref{eq:maxdualfunction}) which is the outer problem. The dual problem can be solved by addressing both problems iteratively, where in each iteration, the optimal power allocation and subchannel allocation can be obtained by using the KKT conditions for a fixed set of Lagrange multipliers, and the outer problem is solved using the (sub)gradient method \cite{boyd2}. \par

Using convex optimization techniques and applying the KKT conditions, the closed-form optimal power allocation on subcarrier $i$ for user $k$ for a given $q_{LR}$ can be obtained as

\begin{equation}
\label{eq:PA}
P_{s,k,i}^{L_{tx}*} = [\frac{q_{LR} + \mu}{ln 2 \Omega_{n}} - \frac{1}{\Gamma_n}]^{+},
\end{equation}

\noindent where $\Omega_{n} =  S_T (1 -\sum_{n \neq k}^{K} \vartheta_{n} L_{s,n} H_{s,n,i}) - \mu $ and $\Gamma_{n} = \frac{L_{s,n} H_{s,n,i}}{\sigma_z}$. Meanwhile, in order to obtain the optimal subchannel allocation $\omega_{s,k,i}^*$, we take the derivative of the subproblem w.r.t. $\omega_{s,k,i}$, which yields

\begin{equation}
\begin{split}
\Theta_{i} &= \frac{\partial \mathcal{L}(\mathbf{P},\bm{\omega},\lambda,\mu,\theta, \gamma)}{\partial \omega_{s,k,i}} \\ & = \Psi - \lambda,
%- \gamma \sum_{j=1}^{N}\omega_{k,j} (E_{s,k,i}^{L_{rx}}+E_{k,j}^{S_{tx}}),
\end{split}
\end{equation}

\noindent where
\begin{equation}
\begin{split}
 & \Psi  = S_{T}(P_{s,k,i}^{L_{tx}}+P_{rx}+P_E \\ &- \sum_{n \neq k} \vartheta_{k} P_{s,k,i}^{L_{tx}} L_{s,n} H_{s,n,i})  -(q_{LR}+\mu) \\ & + \left(1+\log_2(1+\frac{P_{s,k,i}^{L_{tx}} L_{s,k} H_{s,k,i}}{\sigma_z^2}) - \frac{P_{s,k,i}^{L_{tx}} L_{s,k} H_{s,k,i}/\ln 2 \sigma_z^2}{1+P_{s,k,i}^{L_{tx}} L_{s,k} H_{s,k,i}/\sigma_z^2} \right).
\end{split}
\end{equation}

Thus, the subchannel allocation is given by

\begin{equation}
\label{SApolicy}
 \omega_{s,k,i}^*=
\begin{cases}
1, &\text{if $i = \arg \max_{d} \Theta_{d}$,}\\
0, &\text{otherwise.}
\end{cases}
\end{equation}

Moreover, to address the Lagrange multiplier the subgradient method with guaranteed convergence can be used which leads to \cite{chang3}

\begin{equation}
\label{dualupdate1}
\mu^{l+1} = \mu^{l} +\epsilon_{\mu} (R_{L,min} - R_{s,k,i}^{L}),
\end{equation}
\begin{equation}
\label{dualupdate2}
\theta^{l+1} = \mu^{l} + \epsilon_{\theta} (P_{s,k,i}^{L_{tx}} - P_{s,max}),
\end{equation}

\noindent where $\lambda^{l+1}$ and $\mu^{l+1}$ are the values of $\lambda$ and $\mu$ at $l+1$ iterations. $\epsilon_{\lambda}$ and $\epsilon_{\mu}$ are the corresponding step sizes. Since the problem (\ref{eq:optimaNew}) is a convex optimization problem, it is guaranteed that the iteration between the outer and the inner problems converges to the primal optimal solution of (\ref{eq:optimaNew}). \par

To summarize the iterative algorithm between inner and master problems, the multiplier updates can be interpreted as the pricing adjustment \cite{ng3}. Particularly, if the demand of the radio resources exceeds the supply, then the gradient method will raise the prices via adjusting the Lagrange multipliers in the next iteration; otherwise, it will reduce the shadow prices until it is not out of limits. By combining the gradient updates and the subchannel allocation criterion, only one subchannel is selected eventually even though time-sharing is considered for solving the transformed problem in (\ref{eq:optimaNew}). The details of the subchannel and power allocation algorithm is presented in Alg. \ref{alg:RRA}. \par

\begin{algorithm}[htb]
\caption{Subchannel and Power Allocation}
\label{alg:RRA}
\begin{algorithmic}[1]
\STATE Initialize $q_{LR}$, $P_{s,k,i}^{L_{tx}}$ and dual variables; \
\medskip
\WHILE{(!Convergence)}
\STATE Solve the problem (\ref{SApolicy}) for a given $q_{LR}$ and obtain subchannel allocation;\
\STATE Solve the problem (\ref{eq:PA}) and obtain power allocation;\
\STATE Update dual variables according to (\ref{dualupdate1}) and (\ref{dualupdate2});\
\ENDWHILE
\RETURN Obtain subchannel and power allocation policies.\
\end{algorithmic}
\end{algorithm}

We have presented the scheme on how to address the minimization of $\mathcal{E}_{LR}(\omega_{s,k,i},P_{s,k,i}^{L_{tx}}) $. The same procedure can be applied to obtain the optimal solution of minimizing $\mathcal{E}_{SR}(\omega_{k,j},P_{k,j}^{S_{tx}})$. Then we are able to obtain the solution set of (\ref{eq:energyOP}) when considering optimal $k$ is selected.

\subsection{User Scheduling Scheme}
\label{Sec3.3}

For the user scheduling problem, the goal is to select one MT to act as IMT when BS is transmitting data segment and as the data transmitter when delivering data to other MTs after receiving from BS. Therefore, with the assumption that subchannel and power allocations have been done, we are aiming to find a MT that can achieve the best energy efficiency performance considering both LR and SR links. When subchannel and power allocations are done, the objective function can be reformed as,

\begin{equation}
\label{eq:totalenergyRef3}
\min \mathcal{E} (\bm{\rho}) = \frac{U_1(\bm{\rho})}{R_1(\bm{\rho})} +  \frac{U_2(\bm{\rho})}{R_2(\bm{\rho})},
\end{equation}

\noindent where
\begin{equation}
\begin{split}
U_1(\bm{\rho}) &=\sum_k^{K}  \rho_k S_T (P_{s,k,i}^{L_{tx}} + P_{B} + P_{rx}+ P_{E} \\ & - \sum_{n,n \neq k} \vartheta_{n} P_{s,k,i}^{L_{tx}} L_{s,n} H_{s,n,i}),
\end{split}
\end{equation}

\begin{equation}
U_2(\bm{\rho})  = \sum_k^{K} \rho_k S_T(P_{k,j}^{S_{tx}}+  P_{rx}+ 2P_{E}),
\end{equation}
\begin{equation}
R_1(\bm{\rho}) = \sum_{k=1}^{K} \rho_k R_{s,k,i}^{L},
\end{equation}
\begin{equation}
R_2(\bm{\rho}) = \sum_{k=1}^{K} \rho_k R_{k,j}^{S}.
\end{equation}

The reformed problem (\ref{eq:totalenergyRef3}) also subjects to constraints in (\ref{eq:energyOPconstraints}). Consequently, we can obtain the user scheduling criteria as,

\begin{equation}
\label{eq:userscheduling}
 \rho_{k}^*=
\begin{cases}
1, &\text{if $k = \arg \max_{a}\Phi_{a}$,}\\
0, &\text{otherwise.}
\end{cases}
\end{equation}

\noindent where

\begin{equation}
\label{eq:userschedulingBenefits}
\Phi_{a} = \frac{U_1({\rho_a})}{R_{1}({\rho_a})} +  \frac{U_2({\rho_a})}{R_{2}({\rho_a})}.
\end{equation}
%\noindent Here we denote $\mathcal{L}$ as the lagrange function of the problem (ref{eq:totalenergyRef3}). Note that during the dual decomposition process, the involved lagrangian multiplier can be achieve optimally through subgradient algorithm with guaranteed convergence \cite{boyd2}.

\subsection{Solution Description}

The proposed solution involves two subproblems, which are resource (subchannel and power) allocation and user scheduling, and they are interconnected hierarchically. The convergence is guaranteed since in the sub-layer, the two transferred problems are linear for user scheduling indicator and convex for the resource allocation, respectively. The solution is illustrated in Alg. \ref{alg:overall}.\par

\begin{algorithm}[htb]
\caption{Solution Description}
\label{alg:overall}
\begin{algorithmic}[1]
\STATE Initialize $q_{LR},q_{SR}, \bm{\rho}$; \
\medskip
\WHILE{(!Convergence)}
\STATE Obtain subchannel and power allocation policies according to Alg. \ref{alg:RRA} for a given $q_{LR},q_{SR}$ and $\bm{\rho}$;\
\STATE Obtain $q_{LR}$ and $q_{SR}$ according to Alg. \ref{alg:IRA}; \
\STATE Update dual variables according to (\ref{dualupdate1}) and (\ref{dualupdate2});\
\STATE Update the $q_{LR},q_{SR}, \bm{\rho}$; \
\ENDWHILE
\RETURN Obtain user scheduling and resource allocation solutions.\
\end{algorithmic}
\end{algorithm}

\section{Simulation Results}
\label{Sec4}

\subsection{Simulation Setting}
We present the performance evaluation in this section. For LR transmission link, the Stanford University SUI-3 channel model is used and modified to include multipath effects \cite{jain} with central frequency is 2GHz. We use the 3-tap channel and signal fading follows Rician distribution. We choose the number of subchannels $N$ to be 64, so the duality gap can be ignored \cite{wyu}. Flat quasi-static fading channels are adopted, hence the channel coefficients are assumed to be constant during a complete data transmission, and can vary one to another independently. For the SR transmission link, the path loss follows the IEEE 802.11ac standards with 5GHz central frequency. We consider the frequency bandwidths on both LR and SR are equal so no extra frequency band is needed. The noise variance is assumed $1$ for simplicity. Although the baseband power $P_E$ and $P_B$ are not constant in general and their values depend on the features of circuit design, in this work it is out of the scope and we assume they are fixed according to \cite{chang1}. The conversion efficiency is assumed as $\vartheta_{k} = 0.5, \forall k$ for simplicity. To illustrate the energy saving performance, we compare our resource allocation scheme with pure multicast transmission, that is, the reference energy consumption is the one when BS use multicast to deliver all data to every MT as the "conventional transmission" shown in Fig. \ref{fig:1}. In the user location setup, we consider BS is about $500$m from MTs and MTs are randomly located in a $50 \times 50 m^2$ square. \par

\subsection{Performance Evaluation}

\begin{figure}[t]
\centering
\includegraphics[height=6cm, width=8cm]{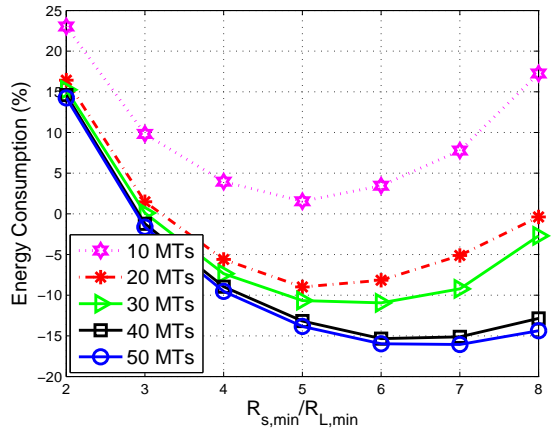}
\caption{User energy consumption of CMC} \label{fig:5}
\end{figure}

\begin{figure}[t]
\centering
\includegraphics[height=6cm, width=8cm]{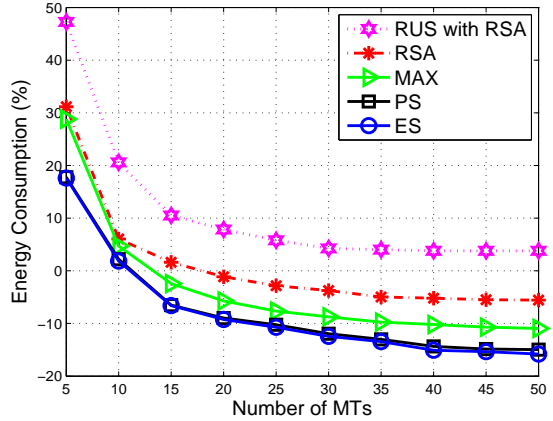}
\caption{User energy consumption, with energy harvesting} \label{fig:2}
\end{figure}

In the following figures, the energy consumption performance of the proposed scheme is examined. First, the energy saving performance of CMC is examined in Fig. \ref{fig:5} where only energy consumption of MTs are shown. For each scheduling interval, the energy consumption ratio (EC) is obtained by normalized the presented system with conventional multicasting scheme, i.e.

\begin{equation}
EC = \frac{E_{k,i,j} - \sum_{n,n \neq k}^{K} Q_{s,n,i}}{\sum_{k}^{K}E_{s,k,i}^{L_{rx}}} \times 100 \%.
\end{equation}

In Fig. \ref{fig:5}, we fix the $R_{L,min}=1 bps/Hz$ and vary the value of $R_{S,min}$ to illustrate energy saving benefits of CMC and the effectiveness of advocating SR for CMC. It can be noticed that at first, when $R_{S,min}$ increases, the energy consumption of CMC is decreased. This is due to the fact that the time durations for transmitting and receiving are reduced so as the consumed energy. Since higher data rate requires higher transmit power consumption, at certain level the energy consumption is increased. For example, for the case that there are 10 MTs inside CMC, the best option for obtaining maximized energy saving is $R_{S,min}/R_{L,min} = 5$. The negative value on the y-axis of Fig. \ref{fig:5} implies that the harvested energy at MTs is higher that the consumed energy, which means that the SWIPT is appreciated for the MTs who are facing energy consumption problems. Moreover, one can find that the CMC consists of more MTs can improve the energy saving potential. \par

\begin{figure}[t]
\centering
\includegraphics[height=6cm, width=8cm]{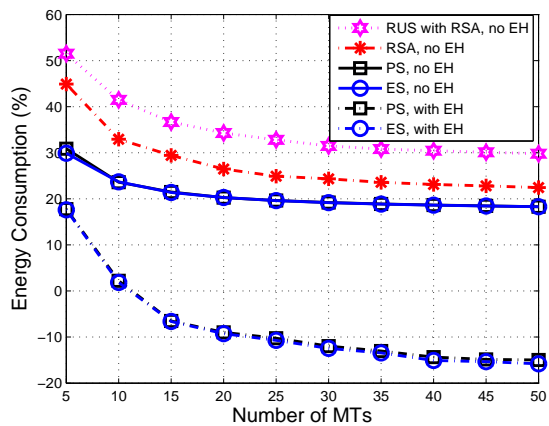}
\caption{User Energy Consumption, with/without energy harvesting comparion} \label{fig:3}
\end{figure}

In Figs \ref{fig:2} and \ref{fig:3}, we assume $R_{S,min}/R_{L,min} = 5$ and present the energy consumption performance of all MTs in a CMC to show the effectiveness of proposed scheme (PS) for the CMC. We compare our proposed scheme (PS) with following reference benchmarks:

\begin{itemize}
  \item the simulation results when random subchannel allocation (RSA) is used instead of the proposed one;
  \item the simulation results when the random user scheduling (RUS) together with RSA is also considered;
  \item the performance when considering a scheduler that can select the MT that has the best channel condition to BS (MAX);
  \item the results obtained by exhaustive searching of IMT and subchannels (ES).
\end{itemize}

In order to see the wireless power transfer impact, we plot the performance of using SWIPT in Fig. \ref{fig:2}. In addition, the comparison of the performance with SWIPT and without SWIPT is illustrated in Fig. \ref{fig:3}. In general, we can see that the CMC is able to reduce the energy consumption in both cases. The energy saving is at least $50 \%$. Even for the cases without energy harvesting, the energy saving for the MTs can be up to $80\%$. When the MTs are able to harvest energy from the RF signal, the energy consumption performance is improved further. In Fig \ref{fig:3}, one can see that enabling wireless power transfer is able to improve the energy consumption performance up to $30\%$, which evidences the significance of SWIPT technique. To summarize the observations, Figs \ref{fig:2} and \ref{fig:3} demonstrate that

\begin{itemize}
  \item the use of CMC is able to reach a promising energy saving gain when comparing with traditional multicasting transmission.
  \item the PS has supervisor performance over RSA, which induces that carefully designing the resource allocation scheme for CMC is necessary.
  \item by recycling energy from RF signal, the energy consumption performance can be improved.
  \item the performance of our proposed scheme is very close to the optimal one.
\end{itemize}

\begin{figure}[t]
\centering
\includegraphics[height=6cm, width=8cm]{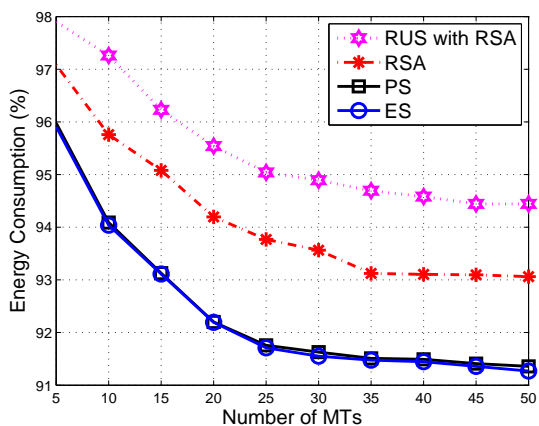}
\caption{System energy consumption,with both BS and user energy consumption} \label{fig:4}
\end{figure}

In Fig. \ref{fig:4}, the energy consumption of the overall system including both BS and MTs is presented. We compare the system performance of PS with the one of RSA and the one of RUS and RSA. The energy consumption ratio is obtained by the energy consumption of proposed scheme normalized by the one when multicasting is invoked, i.e.,

\begin{equation}
EC = \frac{E_{s}^{L_{tx}} + E_{k,i,j} - \sum_{n,n \neq k}^{K} Q_{s,n,i}}{\sum_{k}^{K}E_{s,k,i}^{L_{rx}}+E_{s,tra}^{L_{tx}}} \times 100 \%.
\end{equation}

\noindent where $E_{s,tra}^{L_{tx}}$ is the traditional multicasting transmit energy consumption for single data segment. When jointly considering energy consumption of BS and MTs, the energy saving is up to $9\%$ compared to multicasting transmission. This is mainly due to the fact that energy consumption of BS is fairly lower than the one of MTs and thus, the proposed scheme has less impact on the energy consumption performance of BS.\par

\section{Conclusion}
\label{con}

In this paper we have investigated the problem of resource allocation and user scheduling for OFDMA networks with collaborative mobile clouds. By assuming that the mobile cloud containing numbers of hybrid information decoding and energy harvesting user equipments, we proposed an algorithm which can noticeably obtain energy efficiency performance. The joint optimization problem was solved by addressing two sub-problems including opportunistic selection of information decoding receiver, subchannels and power allocations with the objective of minimizing the energy consumption. It manifested that by designing a proper resource allocation scheme with user scheduling, it is possible to achieve a noticeable gain in the energy consumption for the considered system during receiving process through simulation studies.

\section*{Appendix A}
\label{AppA}
\subsection*{Proof of Theorem 1}

To start with, recalling the energy consumption model

%\begin{equation}
%\begin{split}
%\label{eq:totalenergyApp}
%\mathcal{E} (\bm{\rho},\bm{\omega}, \mathbf{P}) = \sum_{k=1}^{K} \rho_k E_{s}^{L_{tx}} + \sum_{k=1}^{K} \sum_{i=1}^{N} \sum_{j=1}^{N}  E_{k,i,j} - \sum_{i=1}^{N} \sum_{n,n \neq k}  \omega_{s,k,i}  Q_{s,n,i},
%\end{split}
%\end{equation}
\begin{equation}
\label{eq:expressionApp}
\begin{cases}
E_{s}^{L_{tx}} &= \frac{(P_{s,k,i}^{L_{tx}} + P_{B}) S_T}{R_{s,k,i}^{L}}\\
 E_{k,i,j} & = \frac{(P_{rx}+ P_{E})S_T}{R_{s,k,i}^{L}}+\frac{(P_{k,j}^{S_{tx}}+ P_{E})
S_T}{R_{k,j}^{S}}+\frac{(P_{rx}+ P_{E})S_T}{R_{k,j}^{S}} \\
Q_{s,n,i} &= \frac{\vartheta_{n} P_{s,k,i}^{L_{tx}} L_{s,n} H_{s,n,i} S_T}{R_{s,k,i}^{L}}
\end{cases}
\end{equation}

To facilitate the the following analysis, we assume that the subchannels $i$ and $j$
are allocated to scheduled MT $k$ optimally for the transmission process, i.e., $\rho_k = \omega_{s,k,i}  = \omega_{k,j} =1$.
Substitute (\ref{eq:expressionApp}) into (\ref{eq:totalenergy}) we can arrive
%DMT to the fact that SR usually has better channel condition than LR, we denote $R_{k,j}^{S} = \alpha R_{s,k,i}^{L}, \alpha > 1$ \cite{chang1}.

\begin{equation}
\begin{split}
\label{eq:totalenergyApp2}
\mathcal{E} (\mathbf{P}) &= \frac{(P_{s,k,i}^{L_{tx}} + P_{B}) S_T}{R_{s,k,i}^{L}}+ \frac{(P_{rx}+ P_{E})S_T}{R_{s,k,i}^{L}}\\& +\frac{(P_{k,j}^{S_{tx}}+ P_{E})
S_T}{R_{k,j}^{S} }+\frac{(P_{rx}+ P_{E})S_T}{R_{k,j}^{S} } \\&-\frac{\sum_{n,n \neq k} \vartheta_{n} P_{s,k,i}^{L_{tx}} L_{s,n} H_{s,n,i} S_T}{R_{s,k,i}^{L}} \\
&=\underbrace{\frac{P_{s,k,i}^{L_{tx}} - \sum_{n,n \neq k} \vartheta_{n} P_{s,k,i}^{L_{tx}} L_{s,n} H_{s,n,i}+C_1}{R_{s,k,i}^{L}}}_{\mathcal{E}_1} \\& + \underbrace{\frac{(P_{k,j}^{S_{tx}}+ C_2)}{R_{k,j}^{S}}}_{\mathcal{E}_2},
\end{split}
\end{equation}

\noindent $C_1 = P_B+P_{rx}+P_E$ and $C_2 = P_{rx}+2P_E$ are constant for the considered model \cite{chang1}. For the sake of simplicity we use $S_T = 1$. Note that the power allocation policy $\mathbf{P}$ is $\mathbf{P} = \{P_{s,k,i}^{L_{tx}},P_{k,j}^{S_{tx}}\}$. We can see $R_{s,k,i}^{L}$ and $R_{k,j}^{S}$ are the concave functions w.r.t to the power allocation $P_{s,k,i}^{L_{tx}}$ and $P_{k,j}^{S_{tx}}$. Therefore, the functions $\mathcal{E}_1$ and $\mathcal{E}_2$ are strictly quasi-convex function for $P_{s,k,i}^{L_{tx}}$ and $P_{k,j}^{S_{tx}}$, respectively \cite{boyd}. Thus, the objective function is quasi-convex function w.r.t. power allocation policy. Further we can prove that the objective function is first monotonically non-increase and then monotonically non-decrease. The proof can be easily obtained by $\frac{\partial \mathcal{E} (\mathbf{P})}{\partial \mathbf{P} } \vert_{\mathbf{P} \to 0 } \leq 0 $ and $\frac{\partial \mathcal{E} (\mathbf{P})}{\partial \mathbf{P} } \vert_{\mathbf{P} \to \infty} > 0 $

\section*{Appendix B}
\label{AppB}
\subsection*{Proof of Theorem 2}

Similar to the previous proof, we assume that that the subchannels $i$ and $j$ are allocated to scheduled MT $k$ optimally for the transmission process, i.e., $\rho_k = \omega_{s,k,i}  = \omega_{k,j} =1$. Suppose $\mathcal{P}$ is the solution set and let $P_0$ be a solution of (\ref{eq:optimalq}), then we have

\begin{equation}
\label{AppB1}
q_{0} = \frac{U_1(P_{0})}{R_1(P_0)} \leq \frac{U_1(P)}{R_1(P)} , \forall P \in \mathcal{P}.
\end{equation}

Consequently, we can arrive

\begin{equation}
\label{AppB2}
 U_1(P) - q_{0} R_1(P) \geq 0 , \forall P \in \mathcal{P},
\end{equation}
\noindent and

\begin{equation}
\label{AppB3}
U_1(P_0) - q_{0} R_1(P_0) = 0 , \forall P \in \mathcal{P}.
\end{equation}

From (\ref{AppB2}) we see that $min\{ U_1(P) - q_{0} R_1(P) \vert P \in \mathcal{P} \} = 0$. From (\ref{AppB3}) we observe that the minimum value is taken when $P = P_0$. Therefore, the necessary condition can be proved. \par

To prove the sufficient condition, let $P_0$ be a solution of (\ref{eq:optimalqCon}), then we have
\begin{equation}
\label{AppB4}
U_1(P) - q_{0} R_1(P) \geq U_1(P_{0}) - q_{0} R_{1}^{L}(P_0) = 0, \forall P \in \mathcal{P}.
\end{equation}

Hence
\begin{equation}
\label{AppB5}
 U_1(P) - q_{0} R_{1}^{L}(P) \geq 0 , \forall P \in \mathcal{P},
\end{equation}

\noindent and

\begin{equation}
\label{AppB6}
U_1(P_0) - q_{0} R_{1}^{L}(P_0) = 0 , \forall P \in \mathcal{P}.
\end{equation}

From (\ref{AppB5}) we have $\frac{U_1(P)} {R_{1}^{L}(P)} \geq q_0$, that $q_0$ is the minimum of (\ref{eq:optimalq}). From (\ref{AppB6}) we have $\frac{U_1(P_0)}{R_{1}^{L}(P_0)} =  q_{0} $ , that is $P_0$ is a solution of (\ref{eq:optimalq}).

\section*{Appendix C}
\label{AppC}
\subsection*{Proof of Convergence of Algorithm \ref{alg:IRA}}

Similar procedure as shown in \cite{Dinkelbach} can be applied to prove the convergence of the Algorithm \ref{alg:IRA}. For simplicity, assuming that

\begin{equation}
f(q^{\prime}) = \min_{\bm{\omega},\bm{P}} U_1(\bm{\omega},\bm{P}) - q^{\prime} R_1(\bm{\omega},\bm{P})
\end{equation}

where $\bm{\omega} = \{\omega_{s,k,i}\}$ and $\bm{P}=\{ P_{s,k,i}^{L_{tx}} \}$. Then assuming $q^{\backprime} > q^{\prime}$ and considering two optimal resource allocation policies, $\{ \bm{\omega}^{\backprime} ,\bm{P}^{\backprime} \} $ and $\{ \bm{\omega}^{\prime} ,\bm{P}^{\prime} \} $ for $f(q^{\backprime})$ and $f(q^{\prime}) $ respectively, we have

\begin{equation}
\begin{split}
f(q^{\backprime}) & = \min_{\bm{\omega},\bm{P}} U_1(\bm{\omega},\bm{P}) - q^{\backprime} R_1(\bm{\omega},\bm{P}) \\
& = U_1(\bm{\omega}^{\backprime} ,\bm{P}^{\backprime} ) - q^{\backprime} R_1(\bm{\omega}^{\backprime} ,\bm{P}^{\backprime} ) \\
& <  U_1(\bm{\omega}^{\prime} ,\bm{P}^{\prime} ) - q^{\backprime} R_1(\bm{\omega}^{\prime} ,\bm{P}^{\prime} ) \\
& \leq U_1(\bm{\omega}^{\prime} ,\bm{P}^{\prime} ) - q^{\prime} R_1(\bm{\omega}^{\prime} ,\bm{P}^{\prime} ) \\
& = f(q^{\prime}).
\end{split}
\end{equation}

Therefore, we can see $f(q)$ is a strong monotonic decreasing function, i.e. $f(q^{\backprime}) < f(q^{\prime})$, if $q^{\backprime} > q^{\prime}$.
Suppose $\{ \bm{\omega}^{l} ,\bm{P}^{l} \}$ is the optimal resource allocation policies in the $l$th iteration and $q^{l} \neq q^{*}$ and $q^{l+1} \neq q^{*}$. We can observe that $q^{l} > 0$ and $q^{l+1} > 0 $. Since in Algorithm \ref{alg:IRA}, we obtain $q^{l+1} =  U_1( \bm{\omega}^{l} ,\bm{P}^{l}) /  R_1(\bm{\omega}^{l} ,\bm{P}^{l})$. Thus one can arrive

\begin{equation}
\begin{split}
f(q^{l}) & = U_1(\bm{\omega}^{l} ,\bm{P}^{l} ) - q^{l} R_1(\bm{\omega}^{l} ,\bm{P}^{l} ) \\
& =  R_1(\bm{\omega}^{l} ,\bm{P}^{n} )(q^{l+1}- q^{l}).  \\
\end{split}
\end{equation}

As $ f(q^{l}) > 0$ and $R_1(\bm{\omega}^{l} ,\bm{P}^{l} )$, we have $q^{l+1} > q^{l}$. Therefore, we can see that as long as the number of iterations is large enough, $F(q^l) \to 0 $ and $F(q^l) $ satisfies the optimality condition as presented in \textbf{Theorem 2}. To this end, the convergence of Algorithm \ref{alg:IRA} can be proved.
%\clearpage

%\begin{figure}[htb]
%\centering
%\includegraphics[height=6cm, width=9cm]{RelayNetworks.eps}
%\caption{Wireless cooperative relay assisted network.}
%\label{relaynetworks}
%\end{figure}
%
%

\end{document}